\documentclass[a4paper]{report}
\usepackage[utf8]{inputenc}
\usepackage[T1]{fontenc}
\usepackage{RJournal}
\usepackage{amsmath,amssymb,array}
\usepackage{booktabs}

\providecommand{\tightlist}{%
  \setlength{\itemsep}{0pt}\setlength{\parskip}{0pt}}

\usepackage{bm}
\usepackage{booktabs}

\begin{document}

%% do not edit, for illustration only
\sectionhead{Contributed research article}
\volume{XX}
\volnumber{YY}
\year{20ZZ}
\month{AAAA}

\begin{article}

% !TeX root = RJwrapper.tex
\title{Bootstrapping Clustered Data in R using \pkg{lmeresampler}}
\author{by Adam Loy and Jenna Korobova}

\maketitle

\abstract{%
Linear mixed-effects models are commonly used to analyze clustered data
structures. There are numerous packages to fit these models in R and
conduct likelihood-based inference. The implementation of
resampling-based procedures for inference are more limited. In this
paper, we introduce the \pkg{lmeresampler} package for bootstrapping
nested linear mixed-effects models fit via \pkg{lme4} or \pkg{nlme}.
Bootstrap estimation allows for bias correction, adjusted standard
errors and confidence intervals for small samples sizes and when
distributional assumptions break down. We will also illustrate how
bootstrap resampling can be used to diagnose this model class. In
addition, \pkg{lmeresampler} makes it easy to construct interval
estimates of functions of model parameters.
}

\hypertarget{introduction}{%
\subsection{Introduction}\label{introduction}}

Clustered (i.e., multilevel or nested) data structures occur in a wide
range of studies and are commonly analyzed using linear mixed-effects
(LME) models, also referred to as hierarchical linear models or
multilevel linear models
\citep{Pinhiero:2000vf, raudenbush, goldstein2011}. In this paper, we
will restrict attention to the Gaussian response LME model for nested
data structures. For group \(i=1, \ldots, g\), this model is expressed
as \begin{equation}\label{eq:lme}
    \underset{(n_i \times 1)}{\bm{y}_i} = \underset{(n_i \times p)}{\bm{X}_i} \ \underset{(p \times 1)}{\bm{\beta}} + \underset{(n_i \times q)}{\bm{Z}_i} \ \underset{(q \times 1)}{\bm{b}_i} + \underset{(n_i \times 1)}{\bm{\varepsilon}_i}
\end{equation} \begin{equation}\label{eq:dsns}
\begin{pmatrix}\bm{\varepsilon}_i\\ \bm{b}_i \end{pmatrix} \sim \mathcal{N} \left(  
\begin{pmatrix}\bm{0}\\ \bm{0} \end{pmatrix},\ \begin{pmatrix}\sigma^2 \bm{I}_{n_i} & \bm{0}\\ \bm{0} & \bm{D} \end{pmatrix}
\right)
\end{equation} where \(\left(\bm{\varepsilon}_i,\ \bm{b}_i \right)\) and
\(\left(\bm{\varepsilon}_j,\ \bm{b}_j \right)\) are independent for
\(i \ne j\). Here, \(\bm{I}_{n_i}\) denotes the \(n_i\)-dimensional
identity matrix and \(\bm{D}\) is a \(q \times q\) positive-definite
covariance matrix.

In R, Model \ref{eq:lme} is often fit using either \CRANpkg{nlme}
\citep{nlme} or \CRANpkg{lme4} \citep{lme4}. Both packages fit Model
\ref{eq:lme} using either maximum likelihood or restricted maximum
likelihood methods. These methods rely on the distributional assumptions
placed on the residual quantities, \(\bm{\varepsilon}_i\) and
\(\bm{b}_i\), as well as large enough sample sizes. If these conditions
are violated, then the inferential procedures may lead to biased
estimates and/or incorrect standard errors. In such cases, bootstrapping
provides an alternative inferential approach that leads to consistent,
bias-corrected parameter estimates, standard errors, and confidence
intervals. In addition, standard errors and confidence intervals for
functions of model parameters are easily calculated using a bootstrap
procedure.

A variety of bootstrap procedures for clustered data and the LME model
have been proposed and investigated, including the cases (nonparametric)
bootstrap, the residual bootstrap, the parametric bootstrap, the random
effect block (REB) bootstrap, and the wild bootstrap
\citep{Morris:2002tj, Carpenter:2003uy, Field:2007vm, Chambers:2013ba, Modugno2015-kd}.
\citet{VanderLeeden:2008} provide a thorough overview of bootstrapping
LME models and note that these procedures are not widely available in R.
Consequently, R users either need to leave the R ecosystem and use a
specialty software such as MLwiN \citep{Rasbash:2012} to run these
procedures, or need to spend time programming these bootstrap
procedures. Recently, some progress has been made in this area.
\citet{Sanchez-Espigares2009-yg} developed a full-featured bootstrapping
framework for LME models in R; however, this package is not available
and there appears to be no active development. The parametric bootstrap
is now available in \pkg{lme4} via the \texttt{bootMer()} function, and
the \texttt{simulateY()} function in \pkg{nlmeU} \citep{galecki2013}
makes it easy to simulate values of the response variable for
\texttt{lme} model objects, though the user is still required to
implement the remainder of the parametric bootstrap. Bootstrap
procedures for specific inferential tasks have also been implemented.
Most notably, the parametric bootstrap has been implemented to carry out
likelihood ratio tests for the presence of variance components in
\CRANpkg{RLRsim} \citep{rlrsim} and to carry out tests for the mean
structure in \CRANpkg{pbkrtest} \citep{pbkrtest}. \CRANpkg{rptR} uses
the parametric bootstrap to estimate repeatabilities for for LME models
\citep{rptR}.

In this paper, we introduce the \CRANpkg{lmeresampler} package which
implements a suite of bootstrap methods for LME models fit using either
\pkg{nlme} or \pkg{lme4}. In the next section, we will clarify some
notation and terminology for LME models. In
\protect\hyperlink{bootstrap-procedures-for-multilevel-data}{Bootstrap
procedures for multilevel data}, we provide an overview of the bootstrap
methods implemented in \pkg{lmeresampler}. We then give an
\protect\hyperlink{overview-of-lmeresampler}{Overview of lmeresampler},
discuss a variety of
\protect\hyperlink{applications-of-the-bootstrap}{Applications of the
bootstrap} for LME models, and show how to
\protect\hyperlink{bootstrapping-in-parallel}{Bootstrapping in parallel}
works in \pkg{lmeresampler}. We conclude with a
\protect\hyperlink{summary}{Summary} and highlight areas for future
development.

\hypertarget{lme-notation-and-terminology}{%
\subsection{LME notation and
terminology}\label{lme-notation-and-terminology}}

In LME models, inference centers around either the marginal or
conditional distribution of \(\bm{y}_i\), depending on whether global or
group-specific questions are of interest. The marginal distribution of
\(y_i\) for all \(i = 1, \ldots, g\) is given by

\begin{equation}\label{eq:marginalmod}
\bm{y}_i \sim \mathcal{N}\left(\bm{X}_i \bm{\beta},\ \bm{V}_i \right),
\end{equation}

\noindent where
\(\bm{V}_i = \bm{Z}_i \bm{DZ}_i^\prime + \sigma^2 \bm{I}_{n_i}\), and
the conditional distribution of \(\bm{y}_i\) given \(\bm{b}_i\) is
defined as

\begin{equation}\label{eq:conditionalmod}
\bm{y}_i | \bm{b}_i \sim \mathcal{N}\left(\bm{X}_i \bm{\beta} + \bm{Z}_i \bm{b}_i, \ \sigma^2 \bm{I}_{n_i} \right).
\end{equation}

\noindent The parameters of these models can be estimated via maximum
likelihood or restricted maximum likelihood.

In our discussion of bootstrapping LME models, it is also important to
realize that multiple ``residuals'' can be defined. Specifically, there
are three general types of residuals
\citep{Haslett2007-kl, Singer2017-sd}. \dfn{Marginal residuals}
correspond to the errors made in the marginal model,
\(\widehat{\bm r}_i = \bm{y}_i - \bm{X}_i \widehat{\bm{\beta}}\).
Similarly, \dfn{conditional residuals} correspond to the errors made in
the conditional model,
\(\widehat{\bm{e}}_i = \bm{y} - \bm{X}_i \widehat{\bm{\beta}} - \bm{Z}_i \widehat{\bm{b}}_i\).
The \dfn{predicted random effects}, \(\widehat{\bm{b}}_i\), are the last
type of residual quantity.

\hypertarget{bootstrap-procedures-for-multilevel-data}{%
\subsection{Bootstrap procedures for multilevel
data}\label{bootstrap-procedures-for-multilevel-data}}

In \pkg{lmeresampler}, we implement five bootstrap procedures for
multilevel data: a cases (i.e., nonparamteric) bootstrap, a residual
bootstrap, a parametric bootstrap, a wild bootstrap, and a block
bootstrap. In this section, we provide an overview of these bootstrap
approaches. Our discussion focuses on two-level models, but procedures
generalize to higher-level models unless otherwise noted.

\hypertarget{the-cases-bootstrap}{%
\subsubsection{The cases bootstrap}\label{the-cases-bootstrap}}

The cases bootstrap is a fully nonparamteric bootstrap that resamples
the clusters in the data set to generate bootstrap resamples. Depending
on the nature of the data, this resampling can be done only for the
top-level cluster, only at the observation-level within a cluster, or at
both levels. The choice of the exact resampling scheme should be
dictated by the way the data were generated, since the cases bootstrap
generates new data by mimicking this process. \cite{VanderLeeden:2008}
provide a cogent explanation of how to select a resampling scheme. To
help ground the idea of resampling, consider a two-level hierarchical
data set where students are organized into schools.

One version of the cases bootstrap is implemented by only resampling the
clusters. This version of the bootstrap is what \citet{Field:2007vm}
term the cluster bootstrap and \citet{goldstein2011} term the
nonparametric bootstrap. We would choose this resampling scheme, for
example, if schools were chosen at random and then all students within
each school were observed. In this case, the bootstrap proceeds as
follows:

\begin{enumerate}
\def\labelenumi{\arabic{enumi}.}
\tightlist
\item
  Draw a random sample, with replacement, of size \(g\) from the
  clusters.
\item
  For each selected cluster, \(k\), extract all of the cases to form the
  bootstrap sample \(\left(\bm{y}^*_k, \bm{X}^*_k, \bm{Z}^*_k \right)\).
  Since entire clusters are sampled, the total sample size may no longer
  be \(n\).
\item
  Refit the model to the bootstrap sample and extract the parameter
  estimates of interest.
\item
  Repeat steps 1--3 \(B\) times.
\end{enumerate}

\noindent An alternative version of the cases bootstrap only resamples
the observations within clusters, which makes sense in our example if
the schools were fixed and students were randomly sampled within
schools.

\begin{enumerate}
\def\labelenumi{\arabic{enumi}.}
\tightlist
\item
  For each cluster \(i=1,\ldots,g\), draw a random sample of the rows
  the data set, with replacement, to form the bootstrap sample
  \(\left(\bm{y}^*_i, \bm{X}^*_i, \bm{Z}^*_i \right)\).
\item
  Refit the model to the bootstrap sample and extract the parameter
  estimates of interest.
\item
  Repeat steps 1--2 \(B\) times.
\end{enumerate}

\noindent A third version of the cases bootstrap resamples both clusters
and cases within clusters. This is what \citet{Field:2007vm} term the
two-state bootstrap. We would choose this resampling scheme if both
schools and students were sampled during the data collection process.

Regardless of which version of the cases bootstrap you choose, it
requires the weakest conditions: it only requires that the hierarchical
structure in the data set is correctly specified.

\hypertarget{the-parametric-bootstrap}{%
\subsubsection{The parametric
bootstrap}\label{the-parametric-bootstrap}}

The parametric bootstrap simulates random effects and error terms from
the fitted distributions to form bootstrap resamples. Consequently, it
requires the strongest conditions. The parametric bootstrap is
implemented through the following steps:

\begin{enumerate}
\def\labelenumi{\arabic{enumi}.}
\tightlist
\item
  Simulate \(g\) error term vectors, \(e_i^*\), of length \(n_i\) from
  \(\mathcal{N}\left(\bm{0},\widehat{\sigma}^2 \bm{I}_{n_i} \right)\).
\item
  Simulate \(g\) random effects vectors, \(b_i^*\), from
  \(\mathcal{N}\left(\bm{0},\widehat{\bm{D}} \right)\).
\item
  Generate bootstrap responses
  \(y^*_i = \bm{X}_i \widehat{\bm{\beta}} + \bm{Z}_i \bm{b}_i^* + \bm{e}_i^*\).
\item
  Refit the model to the bootstrap responses and extract the parameter
  estimates of interest.
\item
  Repeat steps 2--4 \(B\) times.
\end{enumerate}

\hypertarget{the-residual-bootstrap}{%
\subsubsection{The residual bootstrap}\label{the-residual-bootstrap}}

The residual bootstrap resamples the residual quantities from the fitted
LME model in order to generate bootstrap resamples. A naive
implementation of this type of bootstrap would draw random samples, with
replacement, from the estimated conditional residuals, and the best
linear unbiased predictors (BLUPS); however, this will consistently
underestimate the variability in the data because the residuals are
shrunken toward zero \citep{Morris:2002tj, Carpenter:2003uy}.
\cite{Carpenter:2003uy} solve this problem by ``reflating'' the residual
quantities so that the empirical covariance matrices match the estimated
covariance matrices prior to resampling:

\begin{enumerate}
\def\labelenumi{\arabic{enumi}.}
\item
  Fit the model and calculate the empirical BLUPs,
  \(\widehat{\bm{b}}_i\), and the predicted conditional residuals,
  \(\widehat{\bm{e}}_i\).
\item
  Mean center each residual quantity and reflate the centered residuals.
  Only the process to reflate the predicted random effects is discussed
  below, but the process is analogous for the conditional residuals.

  \begin{enumerate}
  \def\labelenumii{\roman{enumii}.}
  \tightlist
  \item
    Arrange the random effects into a \(g \times q\) matrix, where each
    row contains the predicted random effects from a single group.
    Denote this matrix as \(\widehat{\bm{U}}\). Define
    \(\bm{\widehat{\Gamma}} = \rm{diag}\left(\widehat{\bm{D}}, \widehat{\bm{D}}, \ldots, \widehat{\bm{D}} \right)\),
    the block diagonal covariance matrix of \(\widehat{\bm{U}}\).
  \item
    Calculate the empirical covariance matrix as
    \(\bm{S} = \widehat{\bm{U}}^\prime \widehat{\bm{U}} / g\).
  \item
    Find a transformation of \(\widehat{\bm{U}}\),
    \(\widehat{\bm{U}}^* = \widehat{\bm{U}} \bm{A}\), such that
    \(\widehat{\bm{U}}^{*\prime} \widehat{\bm{U}}^* / g = \bm{\widehat{\Gamma}}\).
    Specifically, we will find \(\bm{A}\) such that
    \(\bm{A}^\prime \widehat{\bm{U}}^\prime \widehat{\bm{U}} \bm{A} / g = \bm{A}^\prime \bm{SA} = \bm{\widehat{\Gamma}}\).
    The choice of \(\bm{A}\) is not unique, so we use the recommendation
    given by \cite{Carpenter:2003uy}:
    \(\bm{A} = \left(\bm{L}_D \bm{L}_S^{-1} \right)^\prime\) where
    \(\bm{L}_D\) and \(\bm{L}_S\) are the Cholesky factors of
    \(\bm{\widehat{\Gamma}}\) and \(\bm{S}\), respectively.
  \end{enumerate}
\item
  Draw a random sample, with replacement, from the set
  \(\lbrace \bm{u}_i^* \rbrace\) of size \(g\), where \(\bm{u}_i^*\) is
  the \(i\)th row of the centered and reflated random effects matrix,
  \(\widehat{\bm{U}}^*\).
\item
  Draw \(g\) random samples, with replacement, of sizes \(n_i\) from the
  set of the centered and reflated conditional residuals,
  \(\lbrace \bm{e}_i^* \rbrace\).
\item
  Generate the bootstrap responses, \(\bm{y}^*_i\), using the fitted
  model equation:
  \(\bm{y}^*_i = \bm{X}_i \widehat{\bm{\beta}} + \bm{Z}_i \widehat{\bm{u}}^*_i + \bm{e}_i^*\).
\item
  Refit the model to the bootstrap responses and extract the parameter
  estimates of interest.
\item
  Repeat steps 3--6 B times.
\end{enumerate}

Notice that the residual bootstrap is a \emph{semiparametric} bootstrap,
since it depends on the model structure (both the mean function and the
covariance structure) but not the distributional conditions
\citep{Morris:2002tj}.

\hypertarget{the-random-effects-block-bootstrap}{%
\subsubsection{The random-effects block
bootstrap}\label{the-random-effects-block-bootstrap}}

Another semiparametric bootstrap is the random effect block (REB)
bootstrap \citep{Chambers:2013ba}. The REB bootstrap can be viewed as a
version of the residual bootstrap where conditional residuals are
resampled from within clusters (i.e., blocks) to allow for weaker
assumptions on the covariance structure of the residuals. The residual
bootstrap requires that the conditional residuals are independent and
identically distributed, whereas the REB bootstrap relaxes this to only
require that the covariance structure of the error terms is similar
across clusters. In addition, the REB bootstrap utilizes the marginal
residuals to calculate nonparametric predicted random effects rather
than relying on the model-based empirical best linear unbiased
predictors (EBLUPS). \cite{Chambers:2013ba} developed three versions of
the REB bootstrap, all of which have been implemented in
\pkg{lmeresampler}. We refer the reader to \cite{Chambers:2013ba} for a
discussion of when each should be used. It's important to note that at
the time of this writing, that the REB bootstrap has only been explored
for use with two-level models.

\hypertarget{reb0}{%
\paragraph{REB/0}\label{reb0}}

The base algorithm for the REB bootstrap (also known as REB/0) is as
follows:

\begin{enumerate}
\def\labelenumi{\arabic{enumi}.}
\tightlist
\item
  Calculate nonparametric residual quantities for the model.

  \begin{enumerate}
  \def\labelenumii{\alph{enumii}.}
  \tightlist
  \item
    Calculate the marginal residuals for each group,
    \(\bm{r}_i = \bm{y}_i - \bm{X}_i \widehat{\bm{\beta}}\).
  \item
    Calculate the nonparametric predicted random effects,
    \(\tilde{\bm{b}}_i = \left( \bm{Z}_i^\prime \bm{Z}_i \right)^{-1} \bm{Z}^\prime_i \bm{r}_i\).
  \item
    Calculate the nonparametric conditional residuals using the
    residuals quantities obtained in the previous two steps,
    \(\bm{\tilde{e}}_i = \bm{r}_i - \bm{Z}_i \bm{\tilde{b}}_i\).
  \end{enumerate}
\item
  Take a random sample, with replacement, of size \(g\) from the set
  \(\lbrace \tilde{\bm{b}}_i \rbrace\). Denote these resampled random
  effects as \(\bm{\tilde{b}}^*_i\).
\item
  Take a random sample, with replacement, of size \(g\) from the cluster
  ids. For each sampled cluster, draw a random sample, with replacement,
  of size \(n_i\) from that cluster's vector of error terms,
  \(\bm{\tilde{e}}_i\).
\item
  Generate bootstrap responses, \(\bm{y}^*_i\), using the fitted model
  equation:
  \(\bm{y}^*_i = \bm{X}_i \widehat{\beta} + \bm{Z}_i \bm{\tilde{b}}^*_i + \bm{\tilde{e}}^*_i\).
\item
  Refit the model to the bootstrap sample and extract the parameter
  estimates of interest.
\item
  Repeat steps 2--5 \(B\) times.
\end{enumerate}

\hypertarget{reb1}{%
\paragraph{REB/1}\label{reb1}}

The first variation of the REB bootstrap (REB/1) zero centers and
reflates the residual quantities prior to resampling in order to satisfy
the conditions for consistency \citep{shao1995jackknife}. This is the
same process outlined in Step 2 of the residual bootstrap outlined
above.

\hypertarget{reb2}{%
\paragraph{REB/2}\label{reb2}}

The second variation of the REB bootstrap (REB/2 or postscaled REB)
addresses two issues: potential non-zero covariances in the joint
bootstrap distribution of the variance components and potential bias in
the parameter estimates. After the REB/0 algorithm is run, the following
post processing is performed:

\hypertarget{uncorrelate-the-variance-components.}{%
\subparagraph{Uncorrelate the variance
components.}\label{uncorrelate-the-variance-components.}}

To uncorrelate the bootstrap estimates of the variance components
produced by REB/0, \cite{Chambers:2013ba} propose the following
procedure:

\begin{enumerate}
\def\labelenumi{\arabic{enumi}.}
\item
  Apply natural logarithms to the bootstrap distribution of each
  variance component and form the following matrices. Note that we use
  \(\nu\) to denote the total number of variance components.

  \begin{itemize}
  \tightlist
  \item
    \(\bm{S}^*\): a \(B \times \nu\) matrix formed by binding the
    columns of these distributions together
  \item
    \(\bm{M}^*\): a \(B \times \nu\) matrix where each column contains
    the column mean from the corresponding column in \(\bm{S}^*\)
  \item
    \(\bm{D}^*\): a \(B \times \nu\) matrix where each column contains
    the column standard deviation from the corresponding column in
    \(\bm{S}^*\)
  \end{itemize}
\item
  Calculate \(\bm{C}^*=\rm{cov} \left(\bm{S}^* \right)\).
\item
  Calculate
  \(\bm{L}^* = \bm{M}^* + \lbrace \left(\bm{S}^* - \bm{M}^*\right)\bm{C}^{*-1/2} \rbrace \circ \bm{D}^*\),
  where \(\circ\) denotes the Hadamard (elementwise) product.
\item
  Exponentiate the elements of \(\bm{L}^*\). The columns of \(\bm{L}^*\)
  are then uncorrelated versions of the bootstrap variance components.
\end{enumerate}

\hypertarget{center-the-bootstrap-estimates-at-the-original-estimate-values.}{%
\subparagraph{Center the bootstrap estimates at the original estimate
values.}\label{center-the-bootstrap-estimates-at-the-original-estimate-values.}}

To correct bias in the estimation of the fixed effects, apply a mean
correction. For each parameter estimate, \(\widehat{\beta}_k\), adjust
the bootstrapped estimates, \(\widehat{\bm{\beta}}_k^*\) as follows:
\(\widehat{\bm{\beta}}_k^{**} = \widehat{\beta}_k + \widehat{\bm{\beta}}_k^* -{\rm avg} \left( \widehat{ \bm{\beta} }_k^* \right)\).
To correct bias in the estimation of the variance components, apply a
ratio correction. For each estimated variance component ,
\(\widehat{\sigma}^2_v\), adjust the uncorrelated bootstrapped
estimates, \(\widehat{\bm{\sigma}}_v^{2*}\) as follows:
\(\widehat{\bm{\sigma}}_{v}^{2**} = \widehat{\bm{\sigma}}_v^{2*} \circ \lbrace \widehat{\sigma}^2_v / {\rm avg} \left( \widehat{\bm{\sigma}}_v^{2*} \right) \rbrace\)

\hypertarget{the-wild-bootstrap}{%
\subsubsection{The wild bootstrap}\label{the-wild-bootstrap}}

The wild bootstrap also relaxes the assumptions made on the error terms
of the model, allowing heteroscedasticity both within and across groups.
The wild bootstrap is well developed for the ordinary regression model
\citep{Liu1988-zw, Flachaire2005-qo, Davidson2008-vq} and
\cite{Modugno2015-kd} adapt it for the nested LME model.

To begin, we can reexpress model \ref{eq:lme} as

\begin{equation}\label{eq:wild}
\bm{y}_i = \bm{X}_i \bm{\beta} + \bm{v}_i, \text{ where } \bm{v}_i = \bm{Z}_i \bm{b}_i + \bm{\varepsilon}_i.
\end{equation}

\noindent The wild bootstrap proceeds as follows:

\begin{enumerate}
\def\labelenumi{\arabic{enumi}.}
\item
  Draw a random sample, \(w_1, w_2, \ldots, w_g\), from an auxillary
  distribution with mean zero and unit variance.
\item
  Generate bootstrap responses using the reexpressed model equation
  \eqref{eq:wild}:
  \(\bm{y}^*_i = \bm{X}_i \widehat{\bm{\beta}} + \tilde{\bm{v}}_i w_j\),
  where \(\tilde{\bm{v}}_i\) is a heteroscedasticity consistent
  covariance matrix estimator. \cite{Modugno2015-kd} suggest using what
  \cite{Flachaire2005-qo} calls \({{\rm HC}_2}\) or \({{\rm HC}_3}\) in
  the regression context:

  \begin{align}
  {{\rm HC}_2}&: \tilde{\bm{v}}_i = {\rm diag} \left( \bm{I}_{n_i} - \bm{H}_i \right)^{-1/2} \circ \bm{r}_i\\
  {{\rm HC}_3}&: \tilde{\bm{v}}_i = {\rm diag} \left( \bm{I}_{n_i} - \bm{H}_i \right) \circ \bm{r}_i,
  \end{align}

  where
  \(\bm{H}_i = \bm{X}_i \left(\bm{X}_i^\prime \bm{X}_i \right)^\prime \bm{X}_i^\prime\),
  the \(i\)th diagonal block of the orthogonal projection matrix, and
  \(\bm{r}_i\) is the vector of marginal residuals for group \(i\).
\item
  Refit the model to the bootstrap sample and extract the parameter
  estimates of interest.
\item
  Repeat steps 1--3 \(B\) times.
\end{enumerate}

\noindent \cite{Modugno2015-kd} consider two two-point auxiliary
distributions, one suggested by \cite{Mammen1993-eo} and the other
suggested by \cite{Liu1988-zw},

\begin{align}
  F_1: w_i &= \begin{cases}
    -(\sqrt{5} - 1)/2 & \text{ with probability } p=(\sqrt{5}+1) / (2 \sqrt{5})\\
    (\sqrt{5} + 1)/2 & \text{ with probability } 1-p\\
  \end{cases}\\
F_2:  w_i &= \begin{cases}
    -1 & \text{ with probability } 0.5\\
    1 & \text{ with probability } 0.5,\\
  \end{cases}
\end{align}

\noindent finding that \(F_1\) was preferred based on a Monte Carlo
study.

\hypertarget{overview-of-lmeresampler}{%
\subsection{Overview of lmeresampler}\label{overview-of-lmeresampler}}

The \pkg{lmeresampler} package implements the five bootstrap procedures
outlined in the previous section for Gaussian response LME models for
nested data structures fit using either \pkg{nlme} \citep{nlme} or
\pkg{lme4}. The workhorse function in \pkg{lmeresampler} is

\begin{Schunk}
\begin{Sinput}
bootstrap(model, .f, type, B, resample = NULL, reb_type = NULL, hccme, aux.dist)
\end{Sinput}
\end{Schunk}

\noindent The four required parameters to \code{bootstrap()} are:

\begin{itemize}
\item
  \texttt{model}, an \texttt{lme} or \texttt{merMod} fitted model
  object.
\item
  \texttt{.f}, a function defining the parameter(s) of interest that
  should be extracted/calculated for each bootstrap iteration.
\item
  \texttt{type}, a character string specifying the type of bootstrap to
  run. Possible values include: \texttt{"parametric"},
  \texttt{"residual"}, \texttt{"reb"}, \texttt{"wild"}, and
  \texttt{"case"}.
\item
  \texttt{B}, the number of bootstrap resamples to generate.
\end{itemize}

There are also four optional parameters: \texttt{resample},
\texttt{reb\_type}, \texttt{hccme}, and \texttt{aux.dist}.

\begin{itemize}
\item
  If the user sets \texttt{type\ =\ "case"}, then they must also set
  \texttt{resample} to specify what cluster resampling scheme to use.
  \texttt{resample} requires a logical vector of length equal to the
  number of levels in the model. A value of \texttt{TRUE} in the \(i\)th
  position indicates that cases/clusters at that level should be
  resampled. For example, to only resample the clusters (i.e., level 2
  units) in a two-level model, the user would specify
  \texttt{resample\ =\ c(FALSE,\ TRUE)}.
\item
  If the user sets \texttt{type\ =\ "reb"}, then they must also set
  \texttt{reb\_type} to indicate which version of the REB bootstrap to
  run. \texttt{reb\_type} accepts the integers \texttt{0}, \texttt{1},
  and \texttt{2} to indicate REB/0, REB/1, and REB/2, respectively.
\item
  If the user sets \texttt{type\ =\ "wild"}, then they must specify both
  \texttt{hccme} and \texttt{aux.dist}. Currently, \texttt{hccme} can be
  set to \texttt{"hc2"} or \texttt{"hc3"} and \texttt{aux.dist} can be
  set to \texttt{"f1"} or \texttt{"f2"}.
\end{itemize}

The \texttt{bootstrap()} function is a generic function that calls
functions for each type of bootstrap. The user can call the specific
bootstrap function directly if preferred. An overview of the specific
bootstrap functions is given in Table \ref{tab:boots}.

\begin{table}
\centering
\begin{tabular}{l l l} \toprule
Bootstrap  & Function name         & Required arguments\\ \midrule
Cases      & \code{case\_bootstrap}       & \code{model, .f, type, B, resample} \\
Residual   & \code{resid\_bootstrap}   & \code{model, .f, type, B} \\
REB        & \code{reb\_bootstrap}       & \code{model, .f, type, B, reb\_type}\\
Wild       & \code{wild\_boostrap}       & \code{model, .f, type, B, hccme, aux.dist} \\
Parametric & \code{parametric\_boostrap} & \code{model, .f, type, B} \\ \bottomrule
\end{tabular}
\caption{Summary of the specific bootstrap functions called by \samp{bootstrap()} and their required arguments.}
\label{tab:boots}
\end{table}

Each of the specific bootstrap functions performs four general steps:

\begin{enumerate}
\def\labelenumi{\arabic{enumi}.}
\item
  \emph{Setup.} Key information (parameter estimates, design matrices,
  etc.) is extracted from the fitted model to eliminate repetitive
  actions during the resampling process.
\item
  \emph{Resampling.} The setup information is passed to an internal
  resampler function to generate the \texttt{B} bootstrap samples.
\item
  \emph{Refitting.} The model is refit for each of the bootstrap samples
  and the specified parameters are extracted/calculated.
\item
  \emph{Clean up.} An internal completion function formats the original
  and bootstrapped quantities to return a list to the user.
\end{enumerate}

\noindent Each function returns an object of class \texttt{lmeresamp},
which is a list with elements outlined in Table \ref{tab:return}.
\texttt{print()}, \texttt{summary()}, \texttt{plot()}, and
\texttt{confint()} methods are available for \texttt{lmeresamp} objects.

\begin{table}[t]
\centering
\begin{tabular}{l p{5in}} \toprule
Element  & Description\\ \midrule
\code{observed}   & values for the original model parameter estimates.  \\ 
\code{model}      & the original fitted model object.   \\
\code{.f}         & the function call defining the parameters of interest.   \\
\code{replicates} & a $B \times p$ tibble containing the bootstrapped quantities. Each column contains a single bootstrap distribution.   \\
\code{stats}      &  a tibble containing the \code{observed}, \code{rep.mean} (bootstrap mean), \code{se} (bootstrap standard error), and \code{bias} values for each parameter.\\
\code{B}          & the number of bootstrap resamples. performed\\
\code{data}       & the original data set.\\
&\\
\code{seed}       &  a vector of randomly generated seeds that are used by the bootstrap.\\
\code{type}       & a character string specifying the type of bootstrap performed\\
\code{call}       & the user's call to the bootstrap. function \\ 
\code{message}    & a list of length \code{B} giving any messages generated during refitting. An entry will be \code{NULL} if no message was generated. \\ 
\code{warning}    & a list of length \code{B} giving any warnings generated during refitting. An entry will be \code{NULL} if no warning was generated.\\ 
\code{error}      & a list of length \code{B} giving any errors encountered during refitting. An entry will be \code{NULL} if no error was encountered.\\ \bottomrule
\end{tabular}
\caption{Summary of the values returned by the bootstrap functions.}
\label{tab:return}
\end{table}

\pkg{lmeresampler} also provides the \texttt{extract\_parameters()}
helper function to extract the fixed effects and variance components
from \texttt{merMod} and \texttt{lme} objects as a named vector.

\hypertarget{applications-of-the-bootstrap}{%
\subsection{Applications of the
bootstrap}\label{applications-of-the-bootstrap}}

\hypertarget{a-two-level-example-jsp-data}{%
\subsubsection{A two-level example: JSP
data}\label{a-two-level-example-jsp-data}}

As a first application of the bootstrap for nested LME models, consider
the junior school project (JSP) data
\citep{goldstein2011, Mortimore1988-xv} that is stored as
\texttt{jsp728} in \pkg{lmeresampler}. The data set is comprised of
measurements taken on 728 elementary school students across 48 schools
in London.

\begin{Schunk}
\begin{Sinput}
library(lmeresampler)
tibble::as_tibble(jsp728)
\end{Sinput}
\begin{Soutput}
#> # A tibble: 728 x 9
#>    mathAge11 mathAge8 gender class     school normAge11 normAge8 schoolMathAge8
#>        <dbl>    <dbl> <fct>  <fct>     <fct>      <dbl>    <dbl>          <dbl>
#>  1        39       36 M      nonmanual 1         1.80     1.55             22.4
#>  2        11       19 F      manual    1        -2.29    -0.980            22.4
#>  3        32       31 F      manual    1        -0.0413   0.638            22.4
#>  4        27       23 F      nonmanual 1        -0.750   -0.460            22.4
#>  5        36       39 F      nonmanual 1         0.743    2.15             22.4
#>  6        33       25 M      manual    1         0.163   -0.182            22.4
#>  7        30       27 M      manual    1        -0.372    0.0724           22.4
#>  8        17       14 M      manual    1        -1.63    -1.52             22.4
#>  9        33       30 M      manual    1         0.163    0.454            22.4
#> 10        20       19 M      manual    1        -1.40    -0.980            22.4
#> # ... with 718 more rows, and 1 more variable: mathAge8c <dbl>
\end{Soutput}
\end{Schunk}

Suppose we wish to fit a model using the math score at age 8, gender,
and the father's social class to describe math scores at age 11,
including a random intercept for school
\citep[see][p. 28--29]{goldstein2011}. This LME model can be fit using
the \texttt{lmer()} function in \pkg{lme4}.

\begin{Schunk}
\begin{Sinput}
library(lme4)
jsp_mod <- lmer(mathAge11 ~ mathAge8 + gender + class + (1 | school), data = jsp728)
\end{Sinput}
\end{Schunk}

To implement the residual bootstrap to estimate the fixed effects, we
can used the \texttt{bootstrap()} function and set
\texttt{type\ =\ "residual"}.

\begin{Schunk}
\begin{Sinput}
(jsp_boot <- bootstrap(jsp_mod, .f = fixef, type = "residual", B = 2000))
\end{Sinput}
\begin{Soutput}
#> Bootstrap type: residual 
#> 
#> Number of resamples: 2000 
#> 
#>             term   observed   rep.mean         se          bias
#> 1    (Intercept) 14.1577509 14.1814482 0.67296443  0.0236972931
#> 2       mathAge8  0.6388895  0.6380166 0.02479513 -0.0008729385
#> 3        genderM -0.3571922 -0.3515385 0.33878196  0.0056537646
#> 4 classnonmanual  0.7200815  0.7061790 0.37392895 -0.0139024377
#> 
#> There were 0 messages, 0 warnings, and 0 errors.
#> 
#> The most commonly occurring message was: NULL
#> 
#> The most commonly occurring warning was: NULL
#> 
#> The most commonly occurring error was: NULL
\end{Soutput}
\end{Schunk}

\noindent We can then calculate normal, percentile, and basic bootstrap
confidence intervals via \texttt{confint()}.

\begin{Schunk}
\begin{Sinput}
confint(jsp_boot)
\end{Sinput}
\begin{Soutput}
#> # A tibble: 12 x 6
#>    term           estimate    lower  upper type  level
#>    <chr>             <dbl>    <dbl>  <dbl> <chr> <dbl>
#>  1 (Intercept)      14.2   12.8     15.5   norm   0.95
#>  2 mathAge8          0.639  0.591    0.688 norm   0.95
#>  3 genderM          -0.357 -1.03     0.301 norm   0.95
#>  4 classnonmanual    0.720  0.00110  1.47  norm   0.95
#>  5 (Intercept)      14.2   12.9     15.5   basic  0.95
#>  6 mathAge8          0.639  0.590    0.690 basic  0.95
#>  7 genderM          -0.357 -1.02     0.300 basic  0.95
#>  8 classnonmanual    0.720  0.0191   1.44  basic  0.95
#>  9 (Intercept)      14.2   12.8     15.5   perc   0.95
#> 10 mathAge8          0.639  0.588    0.688 perc   0.95
#> 11 genderM          -0.357 -1.01     0.303 perc   0.95
#> 12 classnonmanual    0.720 -0.00191  1.42  perc   0.95
\end{Soutput}
\end{Schunk}

The default setting is to calculate all three intervals, but this can be
restricted by setting the \texttt{type} parameter to \texttt{"norm"},
\texttt{"basic"}, or \texttt{"perc"}.

\hypertarget{simulation-based-model-diagnostics}{%
\subsubsection{Simulation-based model
diagnostics}\label{simulation-based-model-diagnostics}}

\cite{Loy2017-fo} propose using the lineup protocol to diagnose LME
models, since artificial structures are often observed in conventional
residual plots for this model class that are not indicative of a model
deficiency.

As, an example, consider the \texttt{Dialyzer} data set provided by
\pkg{nlme}. The data set is from a study characterizing the water
transportation characteristics of 20 high flux membrane dialyzers, which
were introduced to reduce the time a patient spends on hemodialysis
\citep{Vonesh:1992us}. The dialyzers were studied in vitro using bovine
blood at flow rates of either 200 or 300 ml/min. The study measured the
the ultrafiltration rate (ml/hr) at at even transmembrane pressures (in
mmHg). \citet[Section 5.2.2]{Pinhiero:2000vf} discuss modeling these
data. In this section, we will explore how to create a lineup of
residual plots to investigate the adequacy of the initial homoscedastic
LME model fit by \citet{Pinhiero:2000vf}.

\begin{Schunk}
\begin{Sinput}
library(nlme)
dialyzer_mod <- lme(
  rate ~ (pressure + I(pressure^2) + I(pressure^3) + I(pressure^4)) * QB, 
  data = Dialyzer, 
  random = ~ pressure + I(pressure^2)
)
\end{Sinput}
\end{Schunk}

\citet{Pinhiero:2000vf} continue by constructing a residual plot of the
conditional residuals plotted against the transmembrane pressure to
explore the adequacy of the fitted model (Figure
\ref{fig:resid_dialyzer}). There appears to be increasing spread of the
conditional residuals, which would indicate that the homoscedastic model
is not sufficient.

\begin{Schunk}
\begin{figure}

{\centering \includegraphics[width=0.4\linewidth]{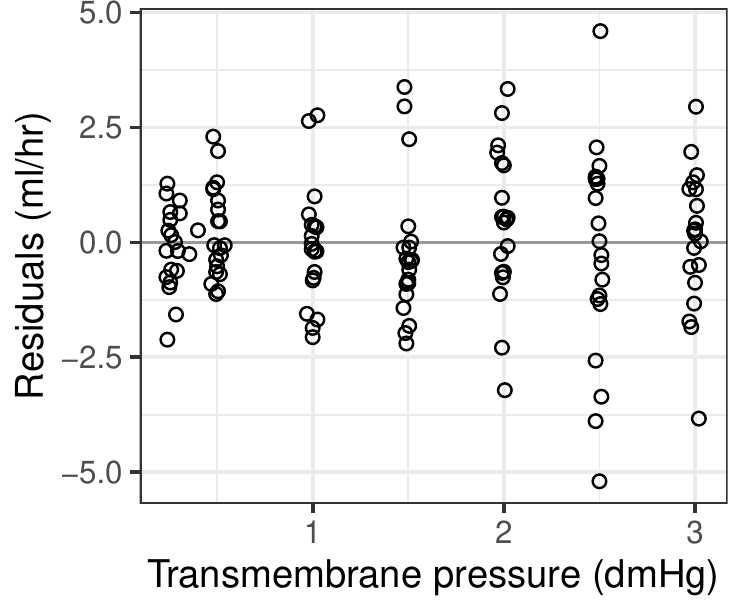} 

}

\caption{\label{fig:resid_dialyzer} Plot of the conditional residuals against the transmembrane pressure for model \code{dialyzer\_mod}.}\label{fig:dialyzer-resid}
\end{figure}
\end{Schunk}

To check if this pattern is actually indicative of a problem, we can
construct a lineup of residual plots. To do this, we must generate data
from a number, say 19, properly specified models to serve as decoy
residual plots. Then, we create a faceted set of residual plots where
the observed residual plot (Figure \ref{fig:resid_dialyzer}) is randomly
assigned to a facet. To generate the residuals from properly specified
models, we use \texttt{bootstrap()} to run a parametric bootstrap, and
use \texttt{hlm\_resid()} from \CRANpkg{HLMdiag} \citep{hlmdiag} to
extract a data frame containing the residuals from each bootstrap
sample:

\begin{Schunk}
\begin{Sinput}
set.seed(1234)
library(HLMdiag)
sim_resids <- bootstrap(dialyzer_mod, .f = hlm_resid, type = "parametric", B = 19)
\end{Sinput}
\end{Schunk}

\noindent The simulated residuals are stored in the \texttt{replicates}
element of the \texttt{sim\_resids} list.
\texttt{sim\_resids\$replicates} is a tibble containing the 19 bootstrap
samples, with the the replicate number stored in the \texttt{.n} column.

\begin{Schunk}
\begin{Sinput}
dplyr::glimpse(sim_resids$replicates)
\end{Sinput}
\begin{Soutput}
#> Rows: 2,660
#> Columns: 15
#> $ id              <dbl> 1, 2, 3, 4, 5, 6, 7, 8, 9, 10, 11, 12, 13, 14, 15, 16,~
#> $ rate            <dbl> -0.8579225, 17.0489071, 34.3658850, 44.8495706, 44.525~
#> $ pressure        <dbl> 0.240, 0.505, 0.995, 1.485, 2.020, 2.495, 2.970, 0.240~
#> $ `I(pressure^2)` <I<dbl>>   0.0576, 0.255025, 0.990025, 2.205225,   4.0804, 6~
#> $ `I(pressure^3)` <I<dbl>>     0.013824,  0.128787625,  0.985074875,  3.274759~
#> $ `I(pressure^4)` <I<dbl>>   0.00331776, 0.065037...., 0.980149...., 4.863017.~
#> $ QB              <fct> 200, 200, 200, 200, 200, 200, 200, 200, 200, 200, 200,~
#> $ Subject         <ord> 1, 1, 1, 1, 1, 1, 1, 2, 2, 2, 2, 2, 2, 2, 3, 3, 3, 3, ~
#> $ .resid          <dbl> -1.19760128, 0.65599523, -0.90104281, 1.39131416, 0.11~
#> $ .fitted         <dbl> 0.3396788, 16.3929119, 35.2669278, 43.4582564, 44.4101~
#> $ .ls.resid       <dbl> -0.51757742, 1.23968773, -1.32231148, 0.59071027, 0.30~
#> $ .ls.fitted      <dbl> -0.3403451, 15.8092194, 35.6881965, 44.2588603, 44.223~
#> $ .mar.resid      <dbl> -2.1236411, -0.3100340, -2.1298265, -0.3453041, -2.455~
#> $ .mar.fitted     <dbl> 1.265719, 17.358941, 36.495712, 45.194875, 46.981067, ~
#> $ .n              <int> 1, 1, 1, 1, 1, 1, 1, 1, 1, 1, 1, 1, 1, 1, 1, 1, 1, 1, ~
\end{Soutput}
\end{Schunk}

Now, we can use the \texttt{lineup()} function from \CRANpkg{nullabor}
\citep{buja2009} to generate the lineup data. \texttt{lineup()} will
randomly insert the observed (\texttt{true}) data into the
\texttt{samples} data, ``encrypt'' the position of the observed data,
and print a message that you can later decrypt in the console.

\begin{Schunk}
\begin{Sinput}
library(nullabor)
lineup_data <- lineup(true = hlm_resid(dialyzer_mod), n = 19, samples = sim_resids$replicates)
\end{Sinput}
\begin{Soutput}
#> decrypt("EnXL zNTN Z6 PJfZTZJ6 Wy")
\end{Soutput}
\begin{Sinput}
dplyr::glimpse(lineup_data)
\end{Sinput}
\begin{Soutput}
#> Rows: 2,800
#> Columns: 15
#> $ id              <dbl> 1, 2, 3, 4, 5, 6, 7, 8, 9, 10, 11, 12, 13, 14, 15, 16,~
#> $ rate            <dbl> -0.8579225, 17.0489071, 34.3658850, 44.8495706, 44.525~
#> $ pressure        <dbl> 0.240, 0.505, 0.995, 1.485, 2.020, 2.495, 2.970, 0.240~
#> $ `I(pressure^2)` <I<dbl>>   0.0576, 0.255025, 0.990025, 2.205225,   4.0804, 6~
#> $ `I(pressure^3)` <I<dbl>>     0.013824,  0.128787625,  0.985074875,  3.274759~
#> $ `I(pressure^4)` <I<dbl>>   0.00331776, 0.065037...., 0.980149...., 4.863017.~
#> $ QB              <fct> 200, 200, 200, 200, 200, 200, 200, 200, 200, 200, 200,~
#> $ Subject         <ord> 1, 1, 1, 1, 1, 1, 1, 2, 2, 2, 2, 2, 2, 2, 3, 3, 3, 3, ~
#> $ .resid          <dbl> -1.19760128, 0.65599523, -0.90104281, 1.39131416, 0.11~
#> $ .fitted         <dbl> 0.3396788, 16.3929119, 35.2669278, 43.4582564, 44.4101~
#> $ .ls.resid       <dbl> -0.51757742, 1.23968773, -1.32231148, 0.59071027, 0.30~
#> $ .ls.fitted      <dbl> -0.3403451, 15.8092194, 35.6881965, 44.2588603, 44.223~
#> $ .mar.resid      <dbl> -2.1236411, -0.3100340, -2.1298265, -0.3453041, -2.455~
#> $ .mar.fitted     <dbl> 1.265719, 17.358941, 36.495712, 45.194875, 46.981067, ~
#> $ .sample         <dbl> 1, 1, 1, 1, 1, 1, 1, 1, 1, 1, 1, 1, 1, 1, 1, 1, 1, 1, ~
\end{Soutput}
\end{Schunk}

\noindent With the lineup data in hand, we can create a lineup of
residual plots using \texttt{facet\_wrap()}:

\begin{Schunk}
\begin{Sinput}
ggplot(lineup_data, aes(x = pressure, y = .resid)) +
  geom_hline(yintercept = 0, color = "gray60") +
  geom_point(shape = 1) +
  facet_wrap(~.sample) +
  theme_bw() +
  labs(x = "Transmembrane pressure (dmHg)", y = "Residuals (ml/hr)")
\end{Sinput}
\begin{figure}

{\centering \includegraphics{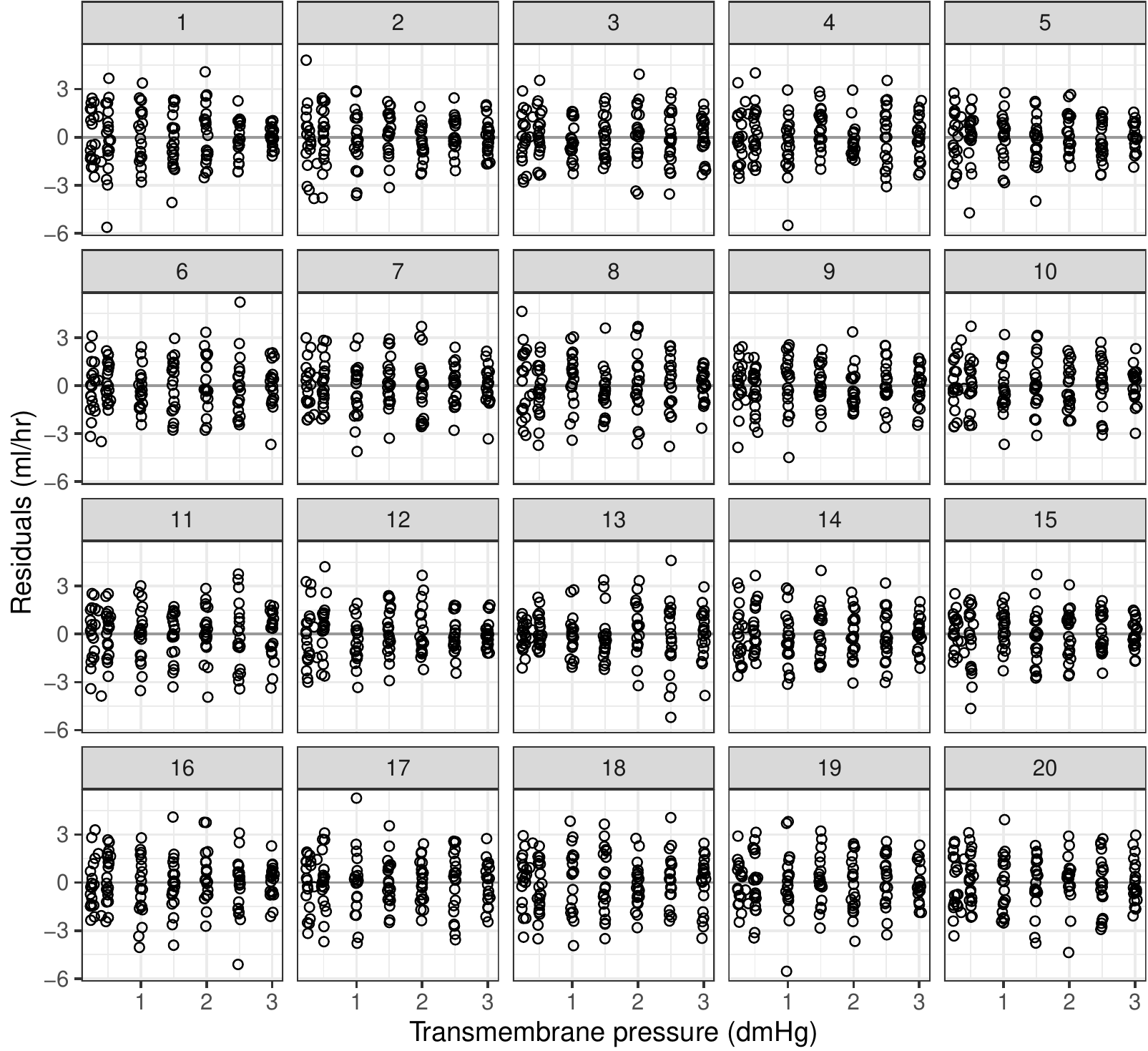} 

}

\caption{\label{fig:lineup} A lineup of the conditional residuals against the transmembrane pressure for model \code{dialyzer\_mod}. One of the facets contains the true residual plot generated from the fitted model, the others are decoys generated using a parametric bootstrap.}\label{fig:dialyzer-lineup}
\end{figure}
\end{Schunk}

\noindent In Figure \ref{fig:lineup}, the observed residual plot is in
position 13. If you can discern this plot from the field of decoys, then
there is evidence that the fitted homogeneous LME model is deficient. In
this case, we believe 13 is discernibly different, as expected based on
the discussion in \citet{Pinhiero:2000vf}. We refer the reader to
\citet{Loy2017-fo} for a discussion of additional diagnostic lineup
plots.

\hypertarget{user-specified-statistics-estimating-repeatibility}{%
\subsubsection{User-specified statistics: Estimating
repeatibility}\label{user-specified-statistics-estimating-repeatibility}}

The beauty of the bootstrap is its flexibility. Interval estimates can
be constructed for functions of model parameters that would otherwise
require more complex derivations. For example, the bootstrap can be used
to estimate the intraclass correlation. The intraclass correlation
mesaures the proportion of the total variance in the response accounted
for by groups, and is an important measure of repeatability in ecology
and evolutionary biology \citep{Nakagawa2010-co}. As a simple example,
we'll consider the \texttt{BeetlesBody} data set in \pkg{rptR}
\citep{rptR}. This simulated data set contains information on body
length (\texttt{BodyL}) and the \texttt{Population} from which the
beetles were sampled. A simple Guassian-response LME model of the form

\[
y_{ij} = \beta_0 + b_i + \varepsilon_{ij}, \qquad b_i \sim \mathcal{N}(0, \sigma^2_b), \qquad \varepsilon_{ij} \sim \mathcal{N}(0, \sigma^2),
\]

\noindent can be used to describe the body length of beetle \(j\) from
population \(i\). The repeatability is then calculated as
\(R = \sigma^2_b / (\sigma^2_b + \sigma^2)\). Below we fit this model
using \texttt{lmer()}:

\begin{Schunk}
\begin{Sinput}
data("BeetlesBody", package = "rptR")
(beetle_mod <- lmer(BodyL ~ (1 | Population), data = BeetlesBody))
\end{Sinput}
\begin{Soutput}
#> Linear mixed model fit by REML ['lmerMod']
#> Formula: BodyL ~ (1 | Population)
#>    Data: BeetlesBody
#> REML criterion at convergence: 3893.268
#> Random effects:
#>  Groups     Name        Std.Dev.
#>  Population (Intercept) 1.173   
#>  Residual               1.798   
#> Number of obs: 960, groups:  Population, 12
#> Fixed Effects:
#> (Intercept)  
#>       14.08
\end{Soutput}
\end{Schunk}

To construct a bootstrap confidence interval for the repeatability, we
first must write a function to calculate it from the fitted model. Below
we write a one-off function for this model to demonstrate a ``typical''
workflow rather than trying to be overly general.

\begin{Schunk}
\begin{Sinput}
repeatability <- function(object) {
  vc <- as.data.frame(VarCorr(object))
  vc$vcov[1] / (sum(vc$vcov))
}
\end{Sinput}
\end{Schunk}

\noindent The original estimate of repeatability can then be quickly
calculated:

\begin{Schunk}
\begin{Sinput}
repeatability(beetle_mod)
\end{Sinput}
\begin{Soutput}
#> [1] 0.2985548
\end{Soutput}
\end{Schunk}

\noindent To construct a bootstrap confidence interval of repeatability,
we run the desired bootstrap procedure, specifying
\texttt{.f\ =\ repeatability} and then pass the results to
\texttt{confint()}.

\begin{Schunk}
\begin{Sinput}
(beetle_boot <- bootstrap(beetle_mod, .f = repeatability, type = "parametric", B = 2000))
\end{Sinput}
\begin{Soutput}
#> Bootstrap type: parametric 
#> 
#> Number of resamples: 2000 
#> 
#>    observed  rep.mean         se        bias
#> 1 0.2985548 0.2845316 0.08997414 -0.01402321
#> 
#> There were 0 messages, 0 warnings, and 0 errors.
#> 
#> The most commonly occurring message was: NULL
#> 
#> The most commonly occurring warning was: NULL
#> 
#> The most commonly occurring error was: NULL
\end{Soutput}
\begin{Sinput}
(beetle_ci <- confint(beetle_boot, type = "basic"))
\end{Sinput}
\begin{Soutput}
#> # A tibble: 1 x 6
#>   term  estimate lower upper type  level
#>   <chr>    <dbl> <dbl> <dbl> <chr> <dbl>
#> 1 ""       0.299 0.136 0.480 basic  0.95
\end{Soutput}
\end{Schunk}

Notice that the \code{term} column of \texttt{beetle\_ci} is an empty
character string since we did not have \code{repeatability} return a
named vector.

Alternatively, we can \texttt{plot()} the results, as shown in Figure
\ref{fig:density}. The plot method for \texttt{lmeresamp} objects uses
\texttt{stat\_halfeye()} from \pkg{ggdist} \citep{ggdist} to render a
density plot with associated 66\% and 95\% percentile intervals.

\begin{Schunk}
\begin{Sinput}
plot(beetle_boot, .width = c(.5, .9)) + 
  labs(
    title = "Bootstrap repeatabilities",
    y = "density",
    x = "repeatability"
  )
\end{Sinput}
\begin{figure}

{\centering \includegraphics[width=0.4\linewidth]{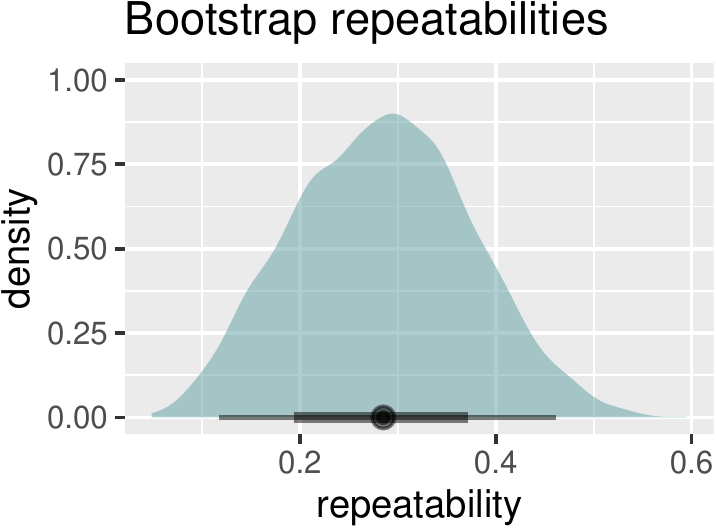} 

}

\caption{\label{fig:density} Density plot with 66\% and 95\% percentile intervals of the bootstrap repeatibilities.}\label{fig:unnamed-chunk-14}
\end{figure}
\end{Schunk}

\hypertarget{bootstrapping-in-parallel}{%
\subsection{Bootstrapping in parallel}\label{bootstrapping-in-parallel}}

Bootstrapping is a computationally demanding task, but bootstrap
iterations do not rely on each other, so they are easy to implement in
parallel. Rather than building parallel processing into
\pkg{lmeresampler}, we have created a utility function,
\texttt{combine\_lmeresamp()}, that allows the user to easily implement
parallel processing using \pkg{doParallel} \citep{doparallel} and
\pkg{foreach} \citep{foreach}. The code is thus concise and simple
enough for users without much experience with parallelization, while
also providing flexibility to the user.

\CRANpkg{doParallel} and \CRANpkg{foreach} default to multicore (i.e.,
forking) functionality when using parallel computing on UNIX operating
systems and snow-like (i.e., clustering) functionality on Windows
systems. In this section, we will use clustering in our example. For
more information on forking, we refer the reader to the vignette in
\citet{doparallel}.

The basic idea behind clustering is to execute tasks as a ``cluster'' of
computers. Each cluster needs to be fed in information separately, and
as a consequence clustering has more overhead than forking. Clusters
also need to be made and stopped with each call to \texttt{foreach()} to
explicitly tell the CPU when to begin and end the parallelization.

Below, we revisit the JSP example and distribute 2000 bootstrap
iterations equally over two cores:

\begin{Schunk}
\begin{Sinput}
library(foreach)
library(doParallel)

set.seed(5678)

# Starting a cluster with 2 cores
no_cores <- 2
cl <- makeCluster(no_cores)
registerDoParallel(cores = no_cores)

# Run 1000 bootstrap iterations on each core
boot_parallel <- foreach(
  B = rep(1000, 2), 
  .combine = combine_lmeresamp,
  .packages = c("lmeresampler", "lme4")
) %dopar% {
  bootstrap(jsp_mod, .f = fixef, type = "parametric", B = B)
}

# Stop the cluster
stopCluster(cl)
\end{Sinput}
\end{Schunk}

\noindent The \texttt{combine\_lmeresamp()} function combines the two
\texttt{lmeresamp} objects that are returned from the two
\texttt{bootstrap()} calls into a single \texttt{lmeresamp} object.
Consequently, working with the returned object proceeds as previously
discussed.

It's important to note that running a process on two cores does not
yield a runtime that is twice as fast as running the same process on one
core. This is because parallelization takes some overhead to split the
processes, so while runtime will substantially improve, it will not
correspond exactly to the number of cores being used. For example, the
runtime for the JSP example run on a single core was

\begin{Schunk}
\begin{Soutput}
#>    user  system elapsed 
#>  38.213   0.368  38.800
\end{Soutput}
\end{Schunk}

\noindent and the runtime for the JSP run on two cores was

\begin{Schunk}
\begin{Soutput}
#>    user  system elapsed 
#>   0.037   0.029  19.324
\end{Soutput}
\end{Schunk}

\noindent These timings were generated using \texttt{system.time()} on a
MacBook Pro with a 2.9 GHz Quad-Core Intel Core i7 processor. In this
set up, running the 2000 bootstrap iterations over two cores reduced the
runtime by a factor of about 2.01, but this will vary for each user.

\hypertarget{summary}{%
\subsection{Summary}\label{summary}}

In this paper, we discussed our implementation of five bootstrap
procedures for nested, Gaussian LME models fit via the \pkg{nlme} or
\pkg{lme4} packages. The \code{bootstrap()} function in
\pkg{lmeresampler} provides a unified interface to these procedures,
allowing users to easily bootstrap their fitted LME models. In our
examples, we illustrated the basic usage of the bootstrap, how it can be
used to create simulation-based visual diagnostics, and how to it can be
used to estimate functions of parameters from the LME model. The
bootstrap approach to inference is computationally intensive, so we have
also demonstrated how users can bootstrap in parallel.

While this paper focused solely on the nested, Gaussian-response LME
model, \pkg{lmeresampler} implements bootstrap procedures for a wide
class of models. Specifically, the cases, residual, and parametric
bootstraps can be used to bootstrap generalized LME models fit via
\code{lme4::glmer()}. Additionally, the parametric bootstrap works with
LME models with crossed random effects, though the results may not be
optimal \citep{Mccullagh2000-st}. Future development of
\pkg{lmeresampler} will focus on implementing additional extensions,
especially for crossed data structures.

\hypertarget{acknowledgements}{%
\subsection{Acknowledgements}\label{acknowledgements}}

We thank Spenser Steele for his contributions to the original code base
of \pkg{lmeresampler}.

\bibliography{loy.bib}

\address{%
Adam Loy\\
Department of Mathematics and Statistics\\%
Carleton College\\ Northfield, MN United States of America\\
\url{https://aloy.rbind.io/}%
\\\textit{ORCiD: \href{https://orcid.org/0000-0002-5780-4611}{0000-0002-5780-4611}}%
\\\href{mailto:aloy@carleton.edu}{\nolinkurl{aloy@carleton.edu}}
}

\address{%
Jenna Korobova\\
Department of Mathematics and Statistics\\%
Carleton College\\ Northfield, MN, United States of America\\
\\\href{mailto:jenna.korobova@gmail.com}{\nolinkurl{jenna.korobova@gmail.com}},
}

\end{article}

\end{document}